\documentclass{article}
\usepackage{frascatiphys,epsfig}
\newcommand{\klpnn}   {\mbox{$K^\circ_L \! \rightarrow \! \pi^\circ \nu \overline{\nu}$ }}
\newcommand{\klpopo}  {\mbox{$K^\circ_L \! \rightarrow \! \pi^\circ \pi^\circ$ }}

\newcommand{\kzpnn}    {\mbox{$K \! \rightarrow \! \pi \nu \overline{\nu}$ }}

\newcommand{\kpp}     {\mbox{$K^+ \! \rightarrow \! \pi^+ \pi^\circ$ }}
\newcommand{\kpen}    {\mbox{$K^+ \! \rightarrow \! \pi^\circ e^+ \nu_e$ }}
\newcommand{\kpnn}    {\mbox{$K^+ \! \rightarrow \! \pi^+ \nu \overline{\nu}$ }}

\newcommand{\bsmix}   {\mbox{$B_s^\circ$---$\overline{B_s^\circ}$ }}
\newcommand{\bsbd}    {\mbox{$\Delta M_{B_d}/\Delta M_{B_s}$ }}
\newcommand{\bpsiks}  {\mbox{$B^\circ_d \! \rightarrow \! \psi K^\circ_S$ }}
\newcommand{\vtd}     {\mbox{$V_{td}$ }}
\newcommand{\Vtd}     {\mbox{$| V_{td} |$ }}

\begin{document}
\title{ EXPERIMENTAL STATUS OF \boldmath \kzpnn }
\author{ S.H. Kettell\\ {\em Brookhaven National Laboratory, Upton, NY 11973-5000 USA} }
\maketitle
\baselineskip=11.6pt
\begin{abstract}

The experimental program for the study of the rare kaon decays,
\kzpnn, is summarized. A review of recent results is provided along
with a discussion of prospects for the future of this program.  The
primary focus of the world-wide kaon program is the two golden modes:
\kpnn and \klpnn.  The first step in an ambitious program to precisely
measure both branching ratios has been successfully completed with the
observation of two \kpnn events by E787. The E949 experiment is poised
to reach an order of magnitude further in sensitivity and to observe
$\sim$10 Standard Model events, and the CKM experiment should observe
$\sim$100 SM events by the end of this decade. Limits on the neutral
analog \klpnn have been set by KTeV and within the next couple of
years will be pushed by E391a.  Measurements of the branching ratio
should be made within the next 10 years by KOPIO, with a goal of $\sim$50
events, and at the JHF, with a goal of up to 1000 events.

\end{abstract}
\baselineskip=14pt
\section{Introduction}

The primary focus in kaon physics today is the two golden modes:
\kpnn\ and \klpnn. These modes are interesting as there is essentially
no theoretical ambiguity in extracting fundamental CKM parameters from
measurements of the branching ratios\cite{sd_kpnn,bb3}.  The intrinsic
theoretical uncertainty in B(\kpnn) is $\sim$7\% and is even smaller
in B(\klpnn), only $\sim$2\%; in both cases the hadronic matrix
element can be extracted from the \kpen\ ($K_{e3}$) decay rate.

The unitarity of the CKM matrix can be expressed as
\[V^*_{us}V_{ud} + V^*_{cs}V_{cd} + V^*_{ts}V_{td}
= \lambda_u + \lambda_c + \lambda_t = 0 \nonumber
\]
with the three vectors $\lambda_i\equiv V^*_{is}V_{id}$ converging to
form a very elongated triangle in the complex plane. The length of
first vector $\lambda_u=V^*_{us}V_{ud}$ is precisely determined from
$K_{e3}$ decay. The length of the third side
$\lambda_t=V^*_{ts}V_{td}$ is measured by \kpnn and the height of the
triangle, $Im\lambda_t$, can be measured by \klpnn.  Branching ratio
measurements of the two \kzpnn modes, along with the well known
$K_{e3}$, will completely determine the unitarity triangle.

Comparison of CKM parameters as measured from the golden \kzpnn,
\bpsiks and \bsbd modes, provide the best opportunity to
over-constrain the unitary triangle and to search for new physics. In
particular, comparisons of
\begin{itemize}
  \item $|\vtd|$ from \kpnn and from the ratio
of the mixing frequencies of $B_d$ and $B_s$ mesons \bsbd~\cite{bb3},
  \item $\beta$ from B(\klpnn)/B(\kpnn) and from the time dependent 
asymmetry in the decay \bpsiks~\cite{sinb,gr-nir}
\end{itemize}
offer outstanding opportunities to explore the Standard Model (SM)
picture of {\it CP}--violation.

The SM prediction for the \kzpnn branching ratios are B(\kpnn) =
$(0.72 \pm 0.21) \times 10^{-10}$ and B(\klpnn) = $(0.26 \pm 0.12)
\times 10^{-10}$~\cite{sd_kpnn}. In addition, an upper limit on
B(\kpnn) can be derived with small uncertainty from the current limit
on $B_s$ mixing; this limit is B(\kpnn) $< 1.32 \times
10^{-10}$~\cite{damb_isidori}.

\section{\boldmath \kpnn}

The Alternating Gradient Synchrotron (AGS) as a high-energy physics
(HEP) research facility has had a broad and rich program in kaon
physics, culminating in the observation of two \kpnn events by E787.
With the recent successful commissioning of the Relativistic Heavy Ion
Collider (RHIC), the primary role of the AGS has shifted to become an
injector of heavy ions for RHIC. Nevertheless, the AGS remains the
highest intensity proton synchrotron in the world and is designed to
be available for $\sim$20 hours/day when not filling RHIC, and as such
retains an important role in the US HEP program.  DOE and BNL have
approved and agreed to fund one new HEP experiment to run at the AGS
between RHIC fills: the E949 experiment seeks to make a precise
measurement of the branching ratio B(\kpnn). While DOE approved E949
to run for 60 weeks, the FY03 budget of the President of the United
States does not include running for E949. At this point the E949
experiment will be terminated after only 12 weeks of running, unless
the US Congress restores funding. The fate of the E949 experiment 
will not be known until Congress initiates and then completes the 
appropriations process between June and October of 2002.

The completion of the Main Injector (MI) at FNAL allows for the simultaneous
operation of a fixed target program along with the Tevatron
collider. The next step in the pursuit of \kpnn will be the CKM experiment,
which plans to use a modest fraction of the MI
protons ($5\times10^{12}$), extracted over a 1-second spill with
minimal bunching of the proton beam, and will push the \kpnn
sensitivity to the current limits of theoretical precision.

\subsection{E787}

\begin{figure}[ht]
\psfig{file=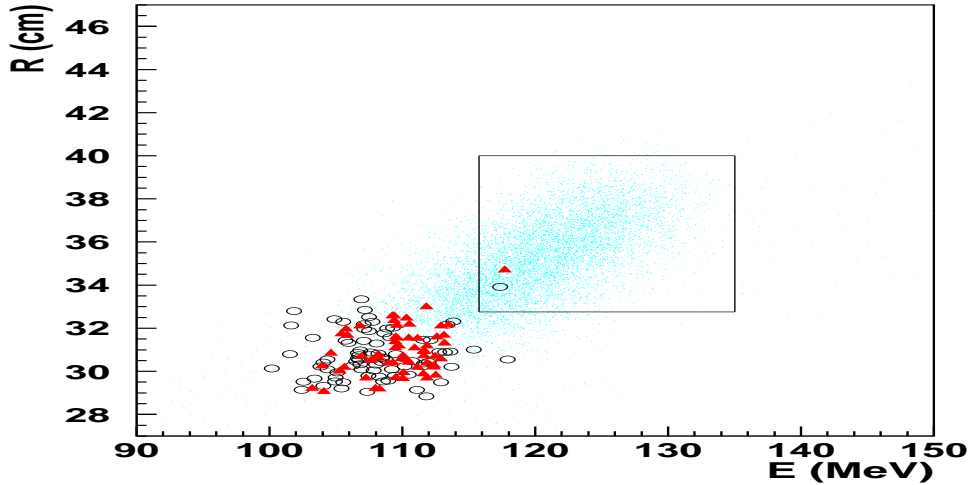,width=5.0in,height=2.5in,angle=0}
\caption{\it Final E787 plot of range vs. energy for events passing all
other cuts. The circles are for 1998 data and the triangles are for
1995--97 data. Two clean \kpnn events are seen in the box. The
remaining events are \kpp background. A \kpnn Monte Carlo sample
(dots) is overlayed on the data.}
\label{fig:pnn}
\end{figure}

The first \kpnn signal was observed in the 1995 data sample of the
E787 experiment~\cite{prl_kpnn_95}. No new events were seen in the
data sample from 1996--97~\cite{prl_kpnn_97}, and with a background of
$0.08\pm0.02$ events and a signal of one event a branching ratio of
B(\kpnn) = $(1.5^{+3.4}_{-1.2})\times 10^{-10}$ was measured. That
event was in fact in a very clean region of the predefined \kpnn
signal region with a SM-signal to background ratio of 35. An analysis of
the final E787 data sample from the 1998-99 run has recently been
reported~\cite{prl_kpnn}. With a measured background of
$0.066^{+0.044}_{-0.025}$, one new event was observed.  The final plot
of range vs. energy from the combined E787 1995--99 data sample for
events passing all other cuts is shown in Figure~\ref{fig:pnn}.
The branching ratio, as determined from these two events, is
\begin{equation}
B(\kpnn) = (1.57^{+1.75}_{-0.82})\times 10^{-10}. 
\end{equation}
This branching ratio is a
factor of two higher than expected in the SM and is higher than
allowed by the current limit on $B_s$ mixing. Of course, the
uncertainty on the BR measurement is large due to limited
statistics and new data from the E949 experiment are eagerly awaited.

\begin{figure}[htb]
\begin{minipage}{0.49\linewidth}
\centerline{\large\bf 1995 Event }
\psfig{file=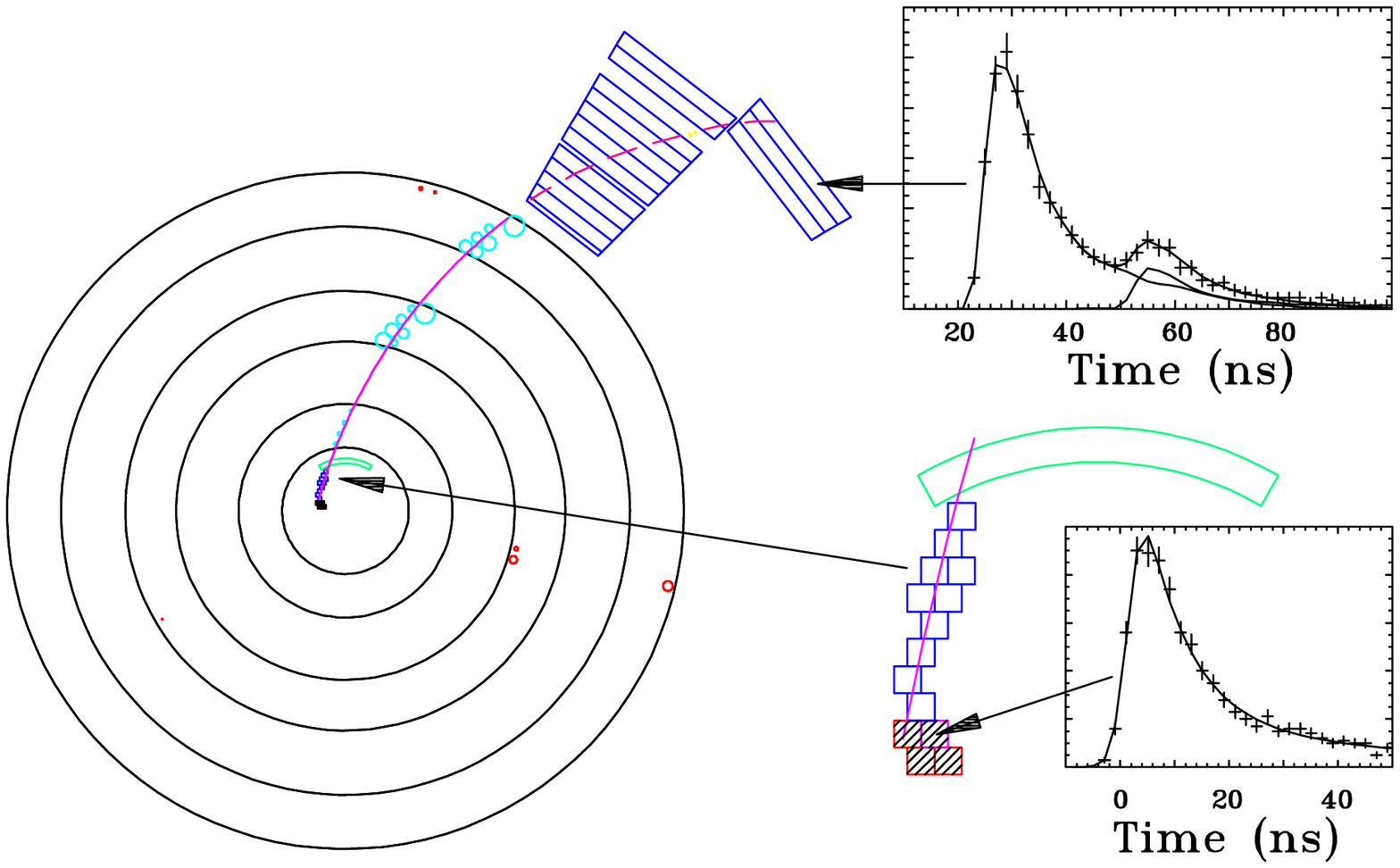,width=2.3in,angle=0}
\end{minipage}\hfill
\begin{minipage}{0.49\linewidth}
\centerline{\large\bf 1998 Event }
\psfig{file=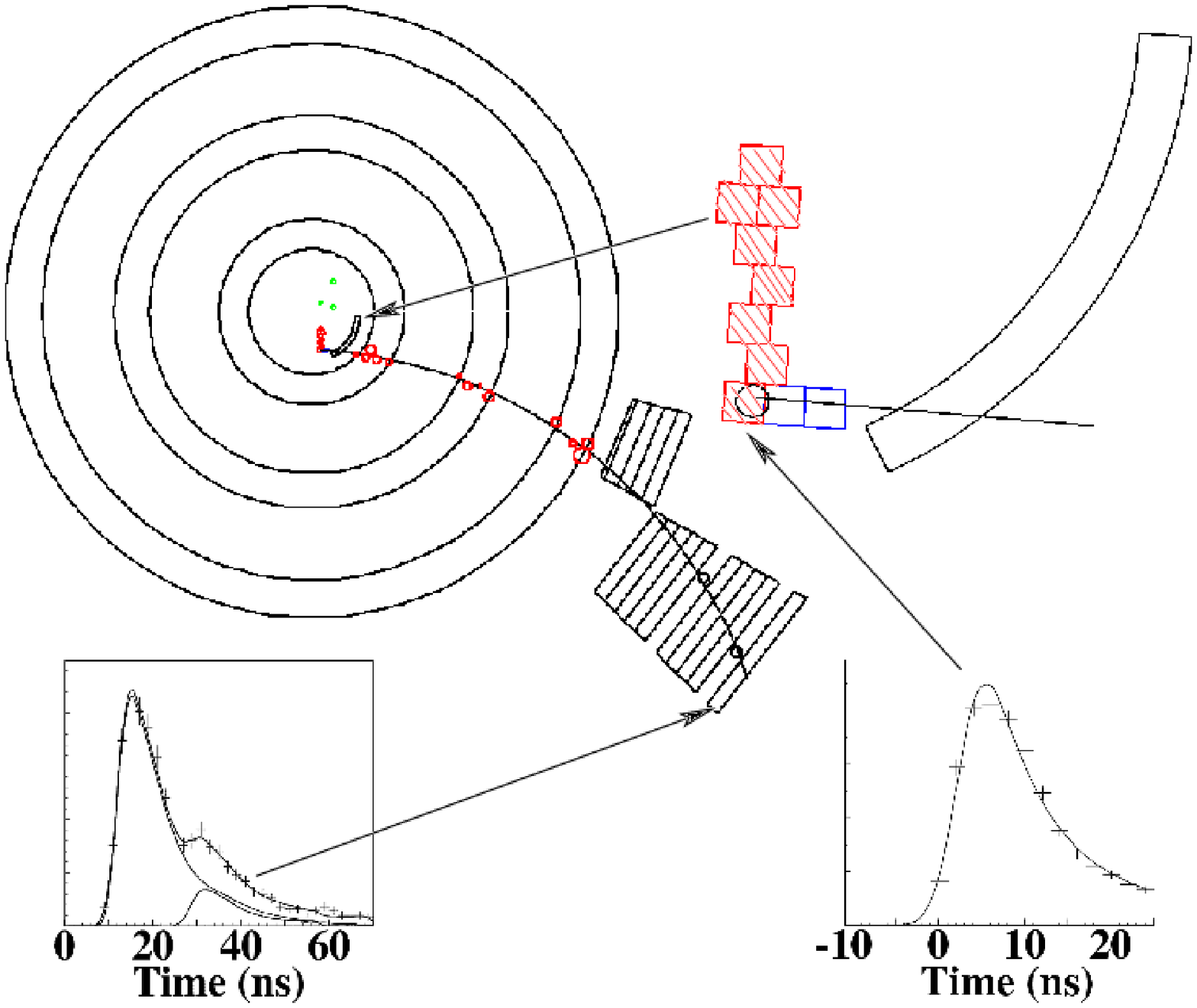,width=2.1in,angle=0}
\end{minipage}\hfill
\caption{Graphical displays of the two \kpnn events, discovered in the
E787 data samples from 1995 and 1998. The display is an end view of the
detector, with  expanded views of the target region, and
with views of the digitized pulses in the fiber where the kaon stopped
and of the $\pi^+\longrightarrow\mu^+$ decay signal in the scintillator where the
pion stopped.}
\label{fig:event}
\end{figure}

The new event found in the 1998 data sample is in a relatively clean
region of the accepted signal region: the SM signal to background
ratio for this event is 3.6. An event display for this event, as well
as the previous event, is shown in Figure~\ref{fig:event}.

Limits on \Vtd and $\lambda_t$ can be obtained (these are 1-$\sigma$ limits
except for $Im\lambda_t$ which is 90\% CL), 
\begin{eqnarray}
0.007< &|V_{td}|& <0.030, \\  \nonumber
2.9 \times 10^{-4} < &|\lambda_t|& <1.2 \times10^{-3}, \\  \nonumber
-0.88 \times 10^{-3} < &Re\lambda_t& < 1.2 \times 10^{-3},\\  \nonumber
&Im\lambda_t& < 1.1 \times10^{-3}.  \nonumber
\end{eqnarray}

\begin{center}
\begin{figure}[htb]
\psfig{file=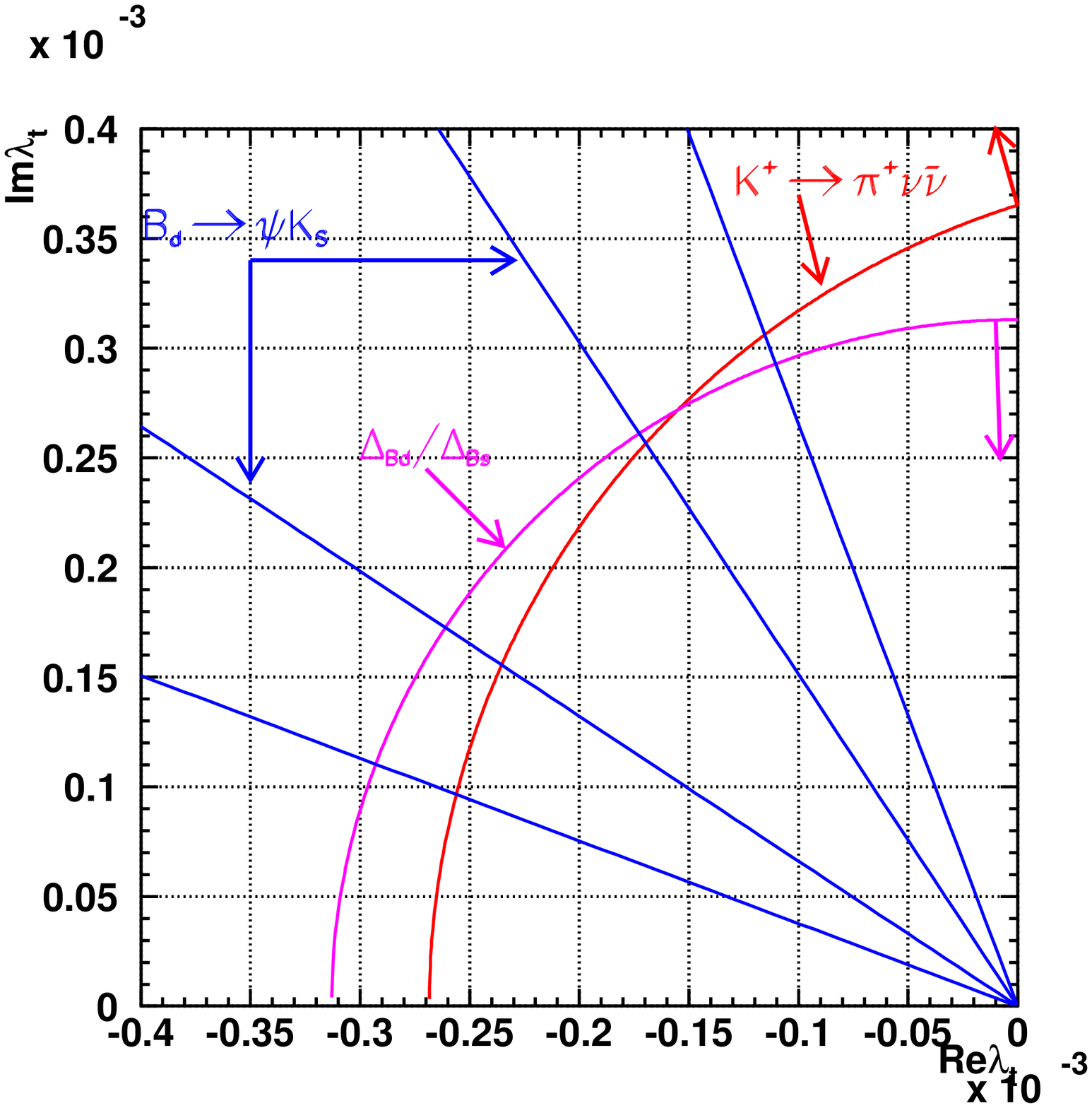,width=3.5in,angle=0}
\caption{Constraints on $\lambda_t$ from the golden modes.
The experimental measurements for \bpsiks and \kpnn
are 90\% CL limits and for \bsbd is a 95\% CL limit.
The theoretical uncertainties in all of these modes are small.  A
measurement of \klpnn will determine  $Im\lambda_t$. (I have used $\Delta M_{B_s}< 14.6 ps^{-1}$, $0.56 < B(\kpnn) < 3.89$ and $\sin(2\beta) = 0.79 \pm 0.13$.)}
\label{fig:rho_eta}
\end{figure}
\end{center}

Even with the large statistical error, this new measurement
provides a non-trivial contribution to global fits of the CKM
parameters~\cite{damb_isidori}.  The constraints on $\lambda_t$ from
this result are shown in Figure~\ref{fig:rho_eta}. The constraints
from the other golden B modes, \bsbd and \bpsiks are shown on the same
plot. One can immediately see that $|\lambda_t|$ is tightly constrained
to a narrow crescent by \kpnn and \bsbd.
New data from the successor to E787, E949, will make a significant
contribution to our knowledge of the CKM parameters.

In addition, E787 has searched for the decay \kpnn in the pion
kinematic region below the \kpp ($K_{\pi2}$)
peak~\cite{prl_kpnn2}. This region contains more of the \kpnn phase
space, but is complicated by a significant background from \kpp decays
with the $\pi^+$ scattering in the scintillating fiber target and
shifting its kinematics into the search region. The data from the
1996 run of E787 has been analyzed ($\sim$20\% of the entire E787 data
sample). One event was observed in the search region, consistent with
the background estimate of $0.73\pm 0.18$. This implies an upper limit
on B(\kpnn) $< 4.2\times 10^{-9}$ (90\% C.L.), and is consistent with
the 2 events observed above the $K_{\pi2}$ peak and the SM spectrum. Some additional
reduction of the background levels in the remaining E787 data may be
possible, but the major focus will shift to the new E949 experiment,
which has significantly enhanced photon veto capabilities that will further
suppress this background. In addition, the next experiment after
E949, CKM at FNAL, will be essentially free of this background
since there is no stopping target.

\subsection{E949}

E949 is an upgraded version of the E787 experiment, planning to
capitalize on the full AGS beam to collect \kpnn data at 14 times the
rate of the E787 run in 1995. The new detector has substantially
upgraded photon veto capabilities, enhanced tracking, triggering,
monitoring, and DAQ capability, and will run at a higher AGS duty
factor and a lower kaon momentum (with an increased fraction of
stopped kaons). It has been designed to reach a sensitivity of
at least 5 times beyond E787 and observe 5--10 SM events. The
background level for E949 measurement of B(\kpnn) above the $K_{\pi2}$
peak is reliably projected from E787 data to be $\sim$10\% of the
Standard Model signal.

\begin{figure}[htb]
\psfig{file=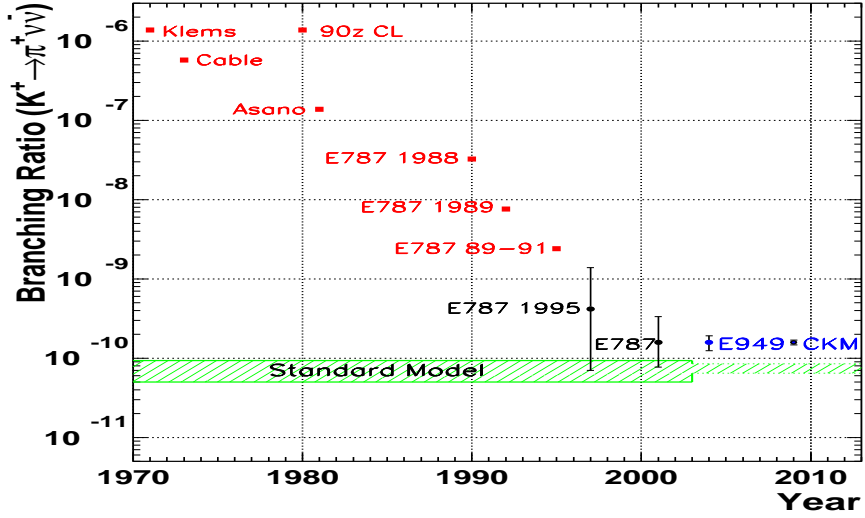,height=3.0in,width=5.0in,angle=0}
\caption{History of the search for \kpnn. The squares represent 90\%
CL limits, the dark circles are the E787 observation of \kpnn, and the
projections of the current central value of the branching ratio to the
proposed E949 and CKM sensitivities. The prediction from the SM is
expected to narrow considerably once \bsmix mixing has been observed.
}
\label{fig:hist}
\end{figure}

E949 should see up to 10 SM events (or 20 events at the branching
ratio measured by E787) within the next couple of years. This is an
exciting opportunity to make a significant contribution to quark
mixing and {\it CP}--violation that should be fully exploited.  A history of
the search for \kpnn is shown in Figure~\ref{fig:hist}.

\subsection{CKM}

The next step towards a precision measurement of B(\kpnn) will be the
CKM experiment at FNAL. CKM has been given scientific (Stage--1)
approval by FNAL and could be running by 2007. CKM plans to use a
novel technique for \kpnn: a decay in flight experiment, with
redundant kinematic constraints from a conventional momentum
spectrometer and a novel velocity spectrometer based on RICH counters.
CKM expects to observe 100 SM signal events in a two year run, using
the Main Injector simultaneously with the Tevatron. The background is
expected to be $\sim$10\% of the SM signal, predominantly from \kpp. A
vacuum of $10^{-6}$ Torr is required to minimize backgrounds from kaon
interactions in the decay volume. CKM will require less than 20\% of
the flux from the Main Injector, but will require a slow extracted
spill of $\sim$1 second duration. CKM will run with a 50 MHz 22 MeV/c beam
with RF separators and about 70\% K$^+$ purity. Figure~\ref{fig:hist} shows a
projected measurement of B(\kpnn) from CKM, assuming the current
central value of the branching ratio.

\section{\boldmath \klpnn}

The decay \klpnn is even cleaner theoretically and is
{\it CP}-violating. However, it is even more challenging
experimentally as all of the particles involved are neutral.

Presently, the best limit on \klpnn is derived in a model-independent
way~\cite{grossman} from the E787 measurement of \kpnn:
\begin{eqnarray}
B(\klpnn) & < & 4.4\times B(\kpnn) \\ \nonumber
         & < & 1.7\times10^{-9} \; \; (90\%\,{\rm CL}).
\label{eq:pnn}
\end{eqnarray}

Of course, it is desirable to observe this mode directly in order
to extract a second constraint on the CKM matrix parameters. 
The current best direct limit is derived from a KTeV search for 
high transverse momentum $\pi^\circ$'s decaying via 
$\pi^\circ\rightarrow e^+e^-\gamma$. From the full 1997 data set,
KTeV observed no events with an expected background of $0.12^{+0.05}_{-0.04}$
and set a 90\%-CL limit~\cite{e799_pnn} of 
\begin{equation}
B(\klpnn) < 5.9\times 10^{-7}.
\end{equation}
Due to the small $\pi^\circ\rightarrow e^+e^-\gamma$ branching ratio
all future experiments plan to use the more copious
$\pi^\circ\rightarrow\gamma\gamma$ mode. KTeV also made a search in
this mode in a special one day test, with a highly collimated `pencil'
beam and observed one background event, most likely from a neutron
interaction, and set a 90\%-CL limit~\cite{e799_pnn_gg} of B(\klpnn)
$< 1.6\times 10^{-6}$.

\subsection{E391}

The next generation of \klpnn experiments will start with E391a at the
High Energy Accelerator Research Organization (KEK)
which hopes to reach a sensitivity of $3\times10^{-10}$. This
experiment will use a technique similar to KTeV, with a pencil beam,
high quality calorimetry and very efficient photon vetos. 
This is the first experiment dedicated to searching for \klpnn and aims to
close the window for non-SM contributions to the decay. It will also serve
as a test bed for the experimental techniques necessary to observe
\klpnn in  future  experiments.
Beam tests
were started in 2001 and the first data-taking run is scheduled for
2003. This experiment plans to eventually move to the Japanese Hadron
Facility (JHF) in $\sim$2007, and attempt to push to a sensitivity of
${\cal O}$(1000) events.

\subsection{KOPIO}

The National Science Board of the National Science Foundation (NSF)
has approved the construction of the two new large experiments at the
BNL AGS: KOPIO and MECO, as components of the Rare Symmetry Violation
Proposal (RSVP). RSVP is planned to be one of the next Major Research
Equipment construction projects at the NSF.

The KOPIO experiment is designed to discover the \klpnn decay and
measure its branching ratio to $\sim$20\%.  KOPIO will make use of a
time-of-flight technique to measure the momentum of the $K_L$ and will
operate at a large targeting angle to improve the $p_K$ resolution and
soften the neutron spectrum to reduce $\pi^\circ$ hadroproduction.
All possible aspects of the decay will be measured: the photon directions
will be measured in a pre-radiator, the times and energy will be precisely
measured in a Shashlyk calorimeter, the time of the kaon's creation is
determined by the proton bunch width from the AGS ($\sim$250 psec).
KOPIO will have a substantial photon veto system and make the same
sort of background measurements as E787, directly from the data, with
two independent tools for attacking the major background, \klpopo,
through both kinematics and photon veto. KOPIO expects to observe 50
SM events, with a background of 50\%. This will allow a determination
of $Im\lambda_t$ to 10\%. KOPIO is expected to start data collection
in $\sim$2006.

\section{Conclusion}

The next decade will be an exciting time for improved understanding of
{\it CP}-violation and quark mixing. It is quite likely that precise
measurements of all four golden modes will be made: \bpsiks at the
B-factories; \bsbd, most likely at the Tevatron; and \kzpnn at BNL, KEK and
FNAL. These measurements will allow a precise determination of CKM parameters
and provide a critical test of the SM
picture of {\it CP}-violation.

%

\section{Acknowledgements}
I would like to thank Takao Inagaki, Bob Tschirhart, Taku Yamanaka,
Takeshi Komatsubara, Shojiro Sugimoto, David Jaffe and Laurie
Littenberg for discussions and/or comments on this paper.  This work was
supported in part under US Department of Energy contract
\#DE-AC02-98CH10886.

%

\end{document}